**Understanding doping, vacancy, lattice stability and superconductivity in $K_xFe_{2-y}Se_2$**

*Yu Liu, Gang Wang\*, Tianping Ying, Xiaofang Lai, Shifeng Jin, Ning Liu, Jiangping Hu and Xiaolong Chen\**

Y. Liu, Prof. G. Wang, T. Ying, X. Lai, S. Jin, N. Liu and Prof. X. Chen
Research & Development Center for Functional Crystals, Beijing National Laboratory for Condensed Matter Physics, Institute of Physics, Chinese Academy of Sciences, Beijing 100190, China
E-mail: gangwang@iphy.ac.cn; chenx29@iphy.ac.cn
Prof. J. Hu
Beijing National Laboratory for Condensed Matter Physics, Institute of Physics, Chinese Academy of Sciences, Beijing 100190, China
Prof. J. Hu and Prof. X. Chen
Collaborative Innovation Center of Quantum Matter, Beijing, China

The discovery of superconductivity in $A_xFe_{2-y}Se_2$ (A = K, Cs, Rb, Tl/Rb, Tl/K) [1-6] triggered another surge of research on iron-based superconductors, which previously were only featured by iron pnictides [7-12] and β-FeSe [13]. The new series of superconductors can be regarded as formation from the intercalation of metals between the FeSe layers of β-FeSe. In comparison to the iron pnictides, metal-intercalated iron selenides are much complicated in terms of structure, chemical composition and phases. The nature of the superconducting (SC) phase, for example, is still in debate though considerable progress has been made over the last a few years [14-16].

One of the most controversial issues is whether the SC phase is Fe-vacancy-free or not, i.e. the FeSe layers in $A_xFe_{2-y}Se_2$ are stoichiometric (y ≈ 0); or off-stoichiometric due to the existence of considerable Fe vacancies. The origin of this issue is largely due to the phase separation that inevitably occurs in these systems at 500 ~ 578 K [17], leading to the coexistence of the insulating phase ($A_2Fe_4Se_5$) and the SC phase [18-23]. The thus-obtained SC phase is not standing freely, instead it intergrows with the insulating phase in the form of nanostrip and its volume fraction is quite low, 10 ~ 20%, as estimated by various measurements [24-29]. This is the main obstacle that prohibits the precise determination of the structure and composition of the SC phase. Ying and coworkers [30, 31] verified that the SC phases in the K-intercalated iron selenides are almost no Fe vacancy in the FeSe layers based on their study of the superconductors obtained via a liquid ammonia route. The neutron diffraction showed that the 43 K SC phase in Li/ammonia co-intercalated FeSe compound is Fe-vacancy-free in a separate study [32]. Guo et al. [33] synthesized a 37 K SC phase containing slight Fe vacancy by the similar liquid ammonia method. In the $K_xFe_{2-y}Se_2$ film grown on $SrTiO_3$ (001) substrate, an Fe-vacancy-free phase $KFe_2Se_2$ with $\sqrt{2} \times \sqrt{5}$ charge ordering was observed in the SC region [34, 35]. Different opinions, however, exist. For example, the SC phases are thought to originate from superstructures due to Fe vacancy, such as $2 \times 4$ [36] or $\sqrt{8} \times \sqrt{10}$ [37]. Meanwhile, phases with disordered Fe vacancies were also reported to be SC [38, 39]. Apart from Fe vacancies, the possibility of the existence of about 20% excess Fe in an SC phase with transition temperature ($T_c$) of 44 K in the K-Fe-Se system was also reported [40].

Another puzzling issue is the existence of several SC phases with different $T_c$ from 30 K to 46 K, which are often mixed in a single sample. For instance, apart from a dominated 30 K phase, a 44 K phase of trace amount sometimes can be observed in a $K_xFe_{2-y}Se_2$ sample obtained by high-temperature route [41]. The amount of 44 K phase can be enhanced in samples with a little less K content (x ≈ 0.6 ~ 0.7), but its real compositions are unknown. The results in ref. [31] revealed that the obtained SC phases differ only in K contents, $K_{0.3}Fe_2Se_2$ with a $T_c$ = 44 K and $K_{0.6}Fe_2Se_2$ with a $T_c$ = 30 K. Superconductivity with $T_c$ of 30 K and 43 K was also observed in $K_xFe_{2-y}Se_2$ samples with $\sqrt{2} \times \sqrt{2} \times 1$ superstructure due to K vacancy ordering, but the phases for different $T_c$ have not been specified [25, 28, 42, 43]. In addition, K-intercalated iron selenide with excess Fe was also proposed to have $T_c$ of 44 K [40].

So far, there has been no consensus on the answers to the above issues. Moreover, the underlying mechanism behind the lattice stability and vacancy are also poorly understood, in particular, the origin for multifarious structures. To address these issues, here, we first study the energetic change and structural evolution as a function of intercalated metal content by taking $K_xFe_{2-y}Se_2$ as an example using first-principles calculations. Two competing factors are found to dominate the formation of the phases and structural evolution. One is due to the energy increase caused by the accumulation of negative charge in FeSe layers. The other is due to the Coulomb attraction between K ion layers and negatively charged FeSe layers. Then we show that the intercalated K content at $0.25 \leq x \leq 0.6$ can stabilize the body centered tetragonal (bct) lattice and Fe is favored to fully occupy its site. At $x > 0.6$, the structure is stabilized by creating Fe vacancies in FeSe layers. At low intercalated K level, $x < 0.25$, the structure will collapse due to the lattice instability. Then a schematic phase diagram is constructed accordingly whereby we speculate the SC phases are $K_xFe_{2-y}Se_2$ with $0.25 \leq x \leq 0.6$ and $y \approx 0$. Our finding sheds some light on understanding this distinct SC family and should be also applicable to other metal-intercalated iron selenides besides $K_xFe_{2-y}Se_2$.

To begin with, we study the formation energy for the process of K intercalation. The process can be expressed as a chemical reaction:

$$xK + 2FeSe \rightarrow K_xFe_2Se_2 \ (0 < x \leq 1)$$

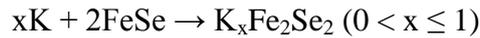

The formation energy per unit cell can be described as $\Delta E_I = E_{KxFe2Se2} - E_{K0Fe2Se2} - xE_K$, where $E_{KxFe2Se2}$ and $E_{K0Fe2Se2}$ are the total energies of $K_xFe_2Se_2$ and an assumed $K_0Fe_2Se_2$ with a similar bct structure but without any K ion, $E_K$ is the energy of elemental K. The variation of $\Delta E_I$ vs. K content is shown in **Figure 1**, which indicates that the intercalation of K into FeSe layers is always energetically favored for the bct structure.

Then we break down $\Delta E_I$ in terms of energy from the structural constituent units and their interactions. For K intercalation process, as shown in Figure 1(a), K atoms lose their valence electrons and form K ion layers between adjacent FeSe layers after K entering into the lattice, while FeSe layers are charged and deformed, and eventually leading to $K_xFe_2Se_2$ with a bct structure. Hence, the following contributions to the total formation energy are considered:

1) The formation of K ion layers: $\Delta E_{K\ ion\ layers} = E_{(K\ ion\ layers)x+} - xE_K$;

2) The electron doping in FeSe layers: $\Delta E_{e-\ doping} = E_{(FeSe\ layers)x-} - E_{FeSe\ layers}$;

3) The deformation of FeSe layers: $\Delta E_{FeSe\ layer\ deformation} = E_{FeSe\ layers} - E_{K0Fe2Se2}$;

4) The Coulomb attraction between K ion layers and FeSe layers: $\Delta E_C = E_{KxFe2Se2} - E_{(FeSe\ layers)x-} - E_{(K\ ion\ layers)x+}$.

And $E_{(K\ ion\ layers)x+}$ is the energy of K ion layers, $E_{(FeSe\ layers)x-}$ the energy of charged FeSe layers, and $E_{FeSe\ layers}$ the energy of FeSe layers in $K_xFe_2Se_2$. Therefore, the total formation energy can be written as the sum of these four energy changes: $\Delta E_I = \Delta E_{K\ ion\ layers} + \Delta E_{e-\ doping} + \Delta E_{FeSe\ layer\ deformation} + \Delta E_C$.

The variations of these four energy contributions as a function of K content are calculated and the results are shown in Figure 1(b), respectively. First of all, FeSe layers are prominent in contributing more and more positive energy to $\Delta E_I$ when they are negatively charged with more and more electrons. This is easily understood since charging a neutral FeSe sheet will incur additional energy like charging a capacitor. Inversely, with the increasing K content, more charges will come into effect in the Coulomb attraction between K ion layers and the charged FeSe layers and thus greatly enhance $|\Delta E_C|$, which will offset sufficiently the increase of energy induced by electron doping into the FeSe layer. In contrast with these two contributions, the other two energy contributions due to the formation of K ion layers and to the deformation of FeSe layers are much smaller. Therefore, we conclude that the former two contributions dominate the energy change during the K intercalation.

To understand the tendency for appearance of Fe vacancy in the K-intercalated iron selenides, we go further to consider the energy change $\Delta E_{Fe\ vacancy}$ (= $\Delta E_I' - \Delta E_I - \mu_{Fe}$, where $\Delta E_I' = E_{KxFe1.94Se2} - E_{K0Fe1.94Se2} - xE_K$ is the formation energy per unit cell in the Fe deficiency structure and $\mu_{Fe}$ is the Fe chemical potential.) by removal of an Fe atom from the unit lattice of $K_xFe_2Se_2$ and its dependence on the K content. It is found that $\Delta E_{Fe\ vacancy}$ only fluctuates above and below zero when $x \leq 0.6$ as shown in **Figure 2**(a). When $x > 0.6$, $\Delta E_{Fe\ vacancy}$ rapidly drops, revealing that Fe vacancies are favored at high levels of K intercalation. This is easily understood since K has a smaller electronegativity than Fe, when the doped electrons due to K is over a limit that Fe-Se bond can accommodate, the surplus electrons will prefer to transfer into $Fe^{2+}$. In this way, Fe is repelled out of the lattice in the form of element. The final compound with Fe vacancy is more energetically favored state. The limit is around at $x = 0.6$. The $K_2Fe_4Se_5$ phase contains 20% Fe vacancy corresponding to $x = 0.8$ is a manifestation of this argument. This tendency of appearance of Fe vacancy with K content is supported by the experimental results [30-32].

Besides the above considerations on the formation energy, the influence of lattice dynamics on the structural stability should also be accounted for. This is, in particular, true for the compounds with low levels of intercalated K, which can induce lattice instability. Figure 2(b) shows the density of phonon states (DOPS) of $K_{0.2}Fe_2Se_2$, where negative frequencies appear, meaning the structure is unstable. Partial DOPS further indicates that it is mainly because of the considerable amount of highly unstable Se atoms, which are unbonded due to the absence of K nearby. The phonon density, however, exhibits no imaginary mode for all other compounds with $x \geq 0.25$. As examples, Figure 2(c) shows the DOPS for $K_{0.25}Fe_2Se_2$ and $K_{0.5}Fe_2Se_2$. The above results demonstrate that the phases of $K_xFe_2Se_2$ free of Fe vacancy are both energetically and dynamically favored in the K intercalation range of $0.25 \leq x \leq 0.6$.

Now we are focused on the trend of structural change upon K intercalation in the range of $0.25 \leq x \leq 0.6$. Both electronic and size effects of K will be taken into accounts. In order to better understand the role of electron doping in the structural change, we charge $K_0Fe_2Se_2$ with various electron concentrations, which allows us to explore the effect of electron doping while excludes the size influence of K. As shown in **Figure 3**(a) and (b), the changes in the lattice constant *a*, the Fe-Se bond and the Se-Fe-Se angle clearly indicate that FeSe layers are stretched along the *ab* plane upon electron doping alone. (Note the results of highly electron-doped are extrapolated from the trend of positively-charged ones simply because the stable negatively-charged $K_0Fe_2Se_2$ cannot always be obtained during the iterative calculations.) For a more realistic case, other factors such as the lattice mismatch and the Coulomb attraction must be included. Figure 3(c) shows the lattice parameter variations with K intercalation, which accounts for all these contributions together with the electron doping. We see that the overall effect of K intercalation, increases the lattice constant *a*, stretches FeSe layers and reduces the lattice constant *c*. For instance, the lattice constant *a* of $K_{0.25}Fe_2Se_2$ and $K_{0.5}Fe_2Se_2$ expands from 3.65 Å to 3.69 Å and *c* shrinks from 14.42 Å to 13.92 Å. The predicted lattice constant *c* of $K_{0.25}Fe_2Se_2$, 14.42 Å, agrees with 14.28(4) Å of $K_{0.3}Fe_2Se_2$ having $T_c$ of 44 K [31]. And the predicted lattice constants *a* and *c* for $K_{0.5}Fe_2Se_2$, 3.69 Å and 13.92 Å, are consistent with the periods observed using scanning tunneling microscope along [110] and [001], 5.5 Å ($\sqrt{2}a$) and 14.1 Å [26], respectively considering the calculation accuracy. It is worth noting that the similar trend of the changes in lattice parameters was observed in other electron-doped $ThCr_2Si_2$ structures, such as $KFe_2As_2$ [9, 44] and $CaFe_2As_2$ [45].

Furthermore, we explore the temperature-dependent stability of structures in the K region of interest. To this end, we perform the molecular dynamics (MD) simulations on $K_{0.25}Fe_2Se_2$ and $K_{0.5}Fe_2Se_2$. Both structures can survive for at least 1 picosecond using $2\sqrt{2} \times 2\sqrt{2} \times 1$ supercells at temperatures up to 500 K, indicating that they are stable at this temperature [46]. It should be worthy to note that the atom displacements of $K_{0.25}Fe_2Se_2$ are larger than those of $K_{0.5}Fe_2Se_2$, suggesting that $K_{0.25}Fe_2Se_2$ is less stable. The variations of free energy with temperature for the two phases are also calculated and shown in Figure 3(d). The free energy is more favorable for $K_{0.5}Fe_2Se_2$ at temperatures above 268 K, consistent with the result obtained by MD. Despite their variations in stability, both of them have the electronic structures similar to $KFe_2Se_2$ (see Figure S3), which suggests that they should have the similar properties to $KFe_2Se_2$. Considering stoichiometry, formation energy, stability and electronic structures, $K_{0.25}Fe_2Se_2$ and $K_{0.5}Fe_2Se_2$ (see Figure S2 for their structures) are proposed to responsible for the observed superconductivity at 44 K and at 30 K [31], respectively. The relative stability difference could also explain the difficulty to obtain the 44 K phase.

Based on the results presented above, we schematize a phase diagram for the $K_xFe_{2-y}Se_2$ system in **Figure 4**. In the K-rich portion with x > 0.6, phases with Fe vacancies tend to exist, agreeing well with the observed antiferromagnetic $K_2Fe_4Se_5$ phase. As for the low level K-intercalated compounds (x < 0.25), there has been no report on the synthesis of free-standing $K_xFe_2Se_2$. We note that the two identified SC phases $K_{0.3}Fe_2Se_2$ and $K_{0.6}Fe_2Se_2$ lie in the region of $0.25 \leq x \leq 0.6$ [31]. $T_c$ is 44 K for the former phase and 30 K for the later one, suggesting that $T_c$ is dependent on K content or the doped electron concentrations. Although experimentally observed K contents in SC phases by far are discrete, recent reports about carrier concentration tuning of $T_c$ from 30 to above 40 K in FeSe thin flakes [47, 48] implies that their variation of $T_c$ with carrier concentration can be continuous, similar to the cases in other high-temperature superconductors. Further improvement of $T_c$ can be expected considering the experimental progresses achieved in single-layer FeSe films [49-54]. Therefore, we infer that the 30 K phase is electron overdoped and the 44 K phase also might not be optimally tuned. Moreover, in alkali-metal-intercalated FeSe compounds prepared at low temperatures, the synergic effect of $NH_3$, $NH_2$, or $C_2H_4(NH_2)_2$ along with alkali metal can stabilize the structures [31, 32, 55]. These offer us an effective strategy to raise $T_c$ of bulk iron-selenide-based superconductors by controlling carrier doping while stabilizing the structures by intercalations or effect of substrates. At the moment, tremendous efforts are needed toward this goal.

Because of the calculation limit and accuracy, we do not consider the slight off-stoichiometric cases, say, the Fe vacancy concentration is less than 3 at.%. It should be pointed out that such slight off-stoichiometry is tolerated in $K_xFe_2Se_2$ just like many other materials, which may be the reason that superconductivity was observed in previous reports of refs. [33, 40].

In conclusion, we carefully investigate the energetic change and structural evolution of $K_xFe_{2-y}Se_2$ as a function of intercalated K content using first-principles calculations. Two factors dominating the formation of the phases and the structural evolution are confirmed. One is due to the accumulation of negative charge in FeSe layers; the other is due to Coulomb attraction between K ion layers and negatively charged FeSe layers. The structural evolution of this series of phases is summarized: at $0.25 \leq x \leq 0.6$, the bct lattice is stable and Fe is favored to full occupy its site; at $x > 0.6$, FeSe layers tend to exclude Fe atoms and create Fe vacancies; and at $x < 0.25$, the structure will collapse for the dynamic instability. A schematic phase diagram is constructed accordingly and the possible route to further improve $T_c$ is suggested. The phases responding to the observed superconductivity are proposed to be $K_{0.25}Fe_2Se_2$ and $K_{0.5}Fe_2Se_2$ in terms of stoichiometry, formation energy, stability and electronic structures. Though based on the study of $K_xFe_{2-y}Se_2$, our results should be meaningful to understand the SC and related phases in metal-intercalated iron selenides and other similar SC systems.

**Acknowledgements**

Y. Liu would like to thank S. J. Shen of Institute of Physics, Chinese Academy of Sciences for the fruitful discussions. This work was partly supported by the National Natural Science Foundation of China (Grant Nos. 51322211, 91422303 and 51532010), the Strategic Priority Research Program (B) of the Chinese Academy of Sciences (Grant No. XDB07020100), Ministry of Education of China (2012 Academic Scholarship Award for

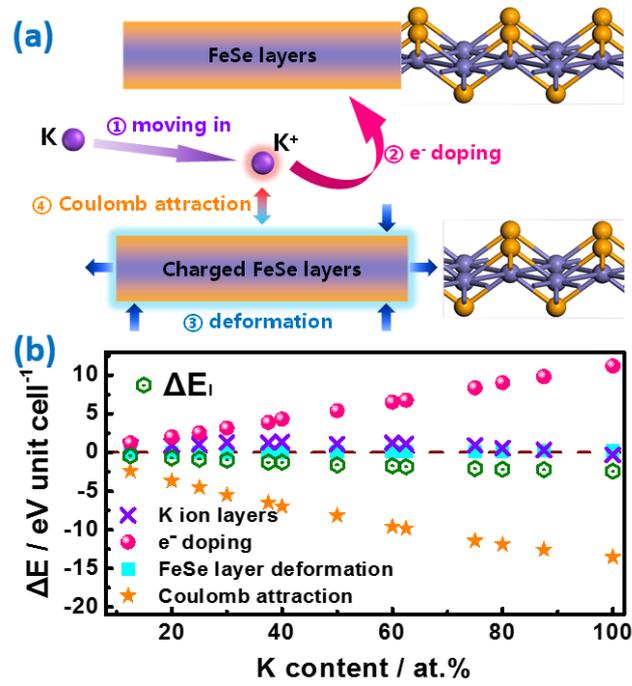

**Figure 1.** (a) The schematic of the K intercalation process. It does not represent a certain structure. (b) The total formation energy and the energy change due to the formation of K ion layers, the electron doping in FeSe layers, the deformation of FeSe layers and the Coulomb attraction between K ion layers and FeSe layers as a function of K content, respectively.

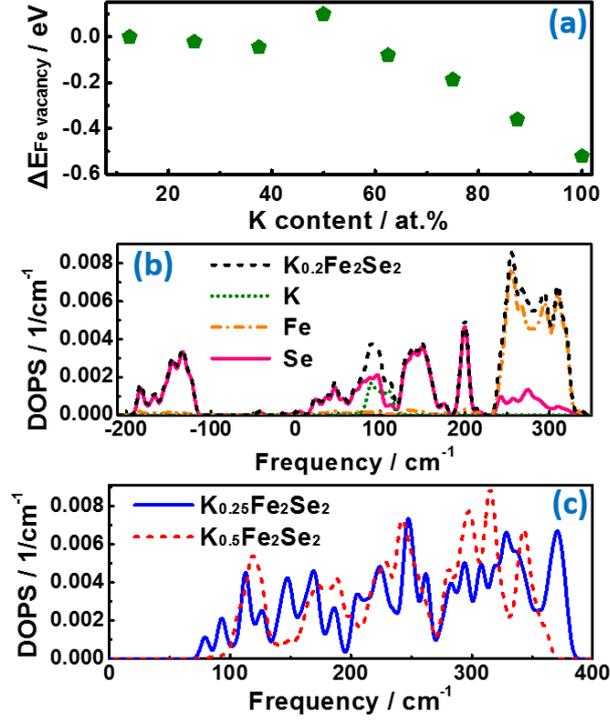

**Figure 2.** (a) The relative variation of formation energy change between $K_xFe_{1.94}Se_2$ and $K_xFe_2Se_2$ as a function of K content. As the absolute value will be determined by Fe chemical potential ($\Delta E_{Fe\ vacancy} = \Delta E_I' - \Delta E_I - \mu_{Fe}$), the change of formation energy between $K_{0.125}Fe_{1.94}Se_2$ and $K_{0.125}Fe_2Se_2$ is set to zero to show the relative values for comparison. (b) Total and partial DOPS of $K_{0.2}Fe_2Se_2$. (c) DOPS of $K_{0.25}Fe_2Se_2$ and $K_{0.5}Fe_2Se_2$.

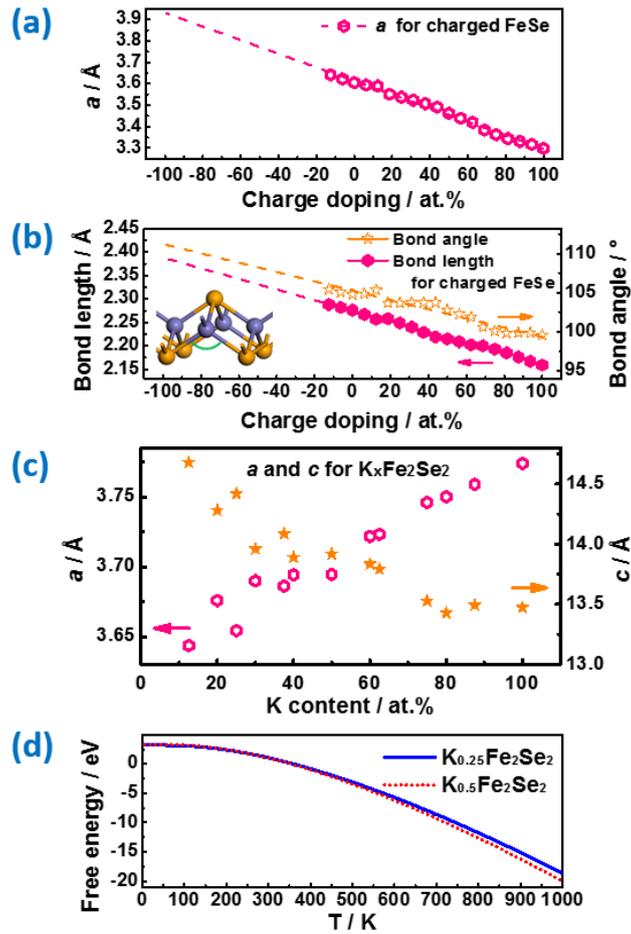

**Figure 3.** The lattice constant $a$ (a), the Fe-Se bond length and the Se-Fe-Se bond angle (see details in the inset) along FeSe layers (b) as a function of charge per unit cell. Positive charge means hole-doped and negative one being electron-doped. Dashed lines represent the results of linear fitting. (c) The variation of the lattice constants $a$ (pink rings) and $c$ (orange stars) with the K content in $K_xFe_2Se_2$. (d) The free energy of $K_{0.25}Fe_2Se_2$ and $K_{0.5}Fe_2Se_2$ as a function of temperature.

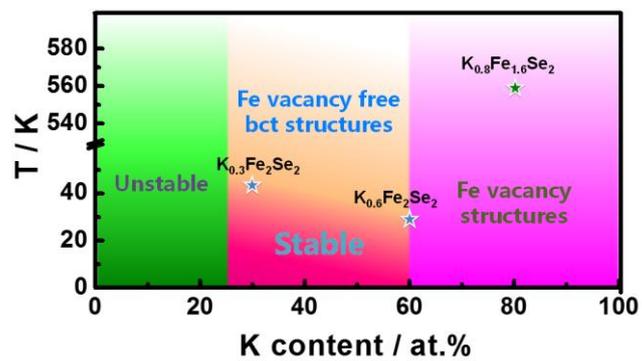

**Figure 4.** Schematic phase diagram of $K_xFe_{2-y}Se_2$.

## Supporting Information

**Understanding doping, vacancy, lattice stability and superconductivity in $K_xFe_{2-y}Se_2$**

*Yu Liu, Gang Wang\*, Tianping Ying, Xiaofang Lai, Shifeng Jin, Ning Liu, Jiangping Hu and Xiaolong Chen\**

Y. Liu, Prof. G. Wang, T. Ying, X. Lai, S. Jin, N. Liu and Prof. X. Chen
Research & Development Center for Functional Crystals, Beijing National Laboratory for Condensed Matter Physics, Institute of Physics, Chinese Academy of Sciences, Beijing 100190, China
E-mail: gangwang@iphy.ac.cn; chenx29@iphy.ac.cn
Prof. J. Hu
Beijing National Laboratory for Condensed Matter Physics, Institute of Physics, Chinese Academy of Sciences, Beijing 100190, China
Prof. J. Hu and Prof. X. Chen
Collaborative Innovation Center of Quantum Matter, Beijing, China

### 1. Simplification of $K_xFe_2Se_2$ superstructures

The example about the minor energy difference of superstructures due to different arrangements of K ions along *c* axis is shown in Figure S1(a) and S1(b). The example about supercells with different lattice constants in *ab* plane have higher $\Delta E_I$ compared to those with the same lattice constants is shown in Figure S1(c) and S1(d).

### 2. The proposed structures of $K_xFe_2Se_2$ (x ≈ 0.3 and 0.6)

The possible structures of $K_xFe_2Se_2$ with uniform K distribution are proposed, $K_{0.25}Fe_2Se_2$ and $K_{0.5}Fe_2Se_2$. For x ≈ 0.3, $K_{0.25}Fe_2Se_2$ with K ion layers in a superstructure of $2 \times 2$ is proposed and the arrangement of K ions is shown in Figure S2(a). In Figure S2(b), $K_xFe_2Se_2$ (x ≈ 0.6) is with the superstructure $\sqrt{2} \times \sqrt{2}$ of K ion layers, which is actually with composition of $K_{0.5}Fe_2Se_2$. This may be the structure mentioned in previous reports [1, 2]. $\Delta E_I$ of $K_{0.25}Fe_2Se_2$ (-0.87 eV/unit cell) is higher than that of $K_{0.5}Fe_2Se_2$ (-1.62 eV/unit cell), implying that $K_{0.25}Fe_2Se_2$ should be less stable than $K_{0.5}Fe_2Se_2$ at the ground state.

## 3. The electronic structures of $K_xFe_2Se_2$ (x ≈ 0.3 and 0.6)

Calculations of total and partial density of states (DOS) are performed to investigate the effects of superstructures on the properties. In Figure S3(a) and S3(b), Fe $3d$ and Se $4p$ orbital hybridization is located between -7 to -3 eV, which is responsible for the formation of FeSe layers in both $K_{0.25}Fe_2Se_2$ and $K_{0.5}Fe_2Se_2$. Part of Fe $3d$ orbital is located around the Fermi level and decides the properties of both structures. As shown in Figure S3(c), the main difference among total DOS of $K_{0.25}Fe_2Se_2$, $K_{0.5}Fe_2Se_2$, and $KFe_2Se_2$ is the locations of their Fermi levels. If the Fermi level of $KFe_2Se_2$ is set to zero, the values of Fermi levels for $K_{0.25}Fe_2Se_2$ and $K_{0.5}Fe_2Se_2$ are about -260 and -200 meV, respectively. So the properties of $K_{0.25}Fe_2Se_2$ and $K_{0.5}Fe_2Se_2$ should be basically the same with that of $KFe_2Se_2$, while their carrier concentrations are different so that they may have different transition temperatures in terms of superconductivity.

## 4. Methods

All the first-principles calculations employ generalized gradient approximation based on density functional theory in the form of Perdew-Burke-Ernzerhof function [3] for exchange-correlation potential, which are performed in the Cambridge Serial Total Energy Package [4]. The self-consistent field method is used with a tolerance of $5.0 \times 10^{-7}$ eV/atom, which is in conjunction with plane-wave basis sets of cutoff energy of 330 eV and ultrasoft pseudopotentials in reciprocal space [5]. According to Monkhorst-Pack special k-point scheme [6], the first Brillouin zone is sampled with grid spacing of 0.04 Å$^{-1}$. Fully optimization of the atomic positions and lattice parameters of all compounds is done until the remanent Hellmann-Feynman forces on all components are less than 0.01 eV/Å. The finite displacement method is used to calculate density of phonon states. In order to examine the dynamic stability of the structures, we perform first-principles finite temperature molecular dynamics simulations with time steps of 10 femtosecond under pressure of 0.1 MPa. Free energies are calculated as a function of temperature using the quasiharmonic approximation based on phonon dispersion [7].

The initial model of $KFe_2Se_2$ is isostructural to $BaFe_2As_2$ [8]. Superstructures due to different arrangements of K ions along *c* axis are not considered for the minor energy difference. For example, the difference of $\Delta E_I$ between two kinds of $1 \times 1 \times 2$ supercells with composition of $K_{0.5}Fe_2Se_2$ [see Figure S1(a) and S1(b)], is only 0.4 meV/unit cell. Superstructures containing one K layer and one FeSe layer due to different arrangements of K ions perpendicular to *c* axis are the main concerns. The $\Delta E_I$ of $4 \times 1 \times 1$ supercell with composition of $K_{0.5}Fe_2Se_2$ [see Figure S1(c)] is 26 meV/unit cell higher than that of $2 \times 2 \times 1$ supercell with the same composition [see Figure S1(d)]. Hence, supercells with different lattice constants in *ab* plane, like $3 \times 1$

× 1, 3 × 2 × 1 and so on, are not discussed in this work for their higher $\Delta E_I$ compared to those with the same lattice constants in *ab* plane. Supercells √2 × √2 × 1, 2 × 2 × 1, √5 × √5 × 1 and 2√2 × 2√2 × 1 are built accordingly with uniform K distribution as the Coulomb repulsion between K ions dominates K distribution. To study the formation of Fe vacancy, a 2√2 × 2√2 × 1 supercell ($a = b > 10.3$ Å and $c > 13.5$ Å), which is large enough to accommodate a single Fe vacancy while keeping the tetragonal symmetry, is therefore chosen. In this case, the Fe content corresponds to y = 0.06. The above structures are only designed to obtain different K contents in $K_xFe_{2-y}Se_2$ (y = 0 or 0.06), which shall not represent the real ones. As K ion ordering cannot guarantee the tetragonal symmetry in the structures, the obtained lattice constants or bond lengths will be the average values. The bond angles are obtained using the average coordinates of Fe and Se atoms. Besides, when we calculate $\Delta E_I$, the total energy of $K_0Fe_2Se_2$ is used instead of that of FeSe so as to offset the additional energy, tens of meV per unit cell induced by the change of cell volume.

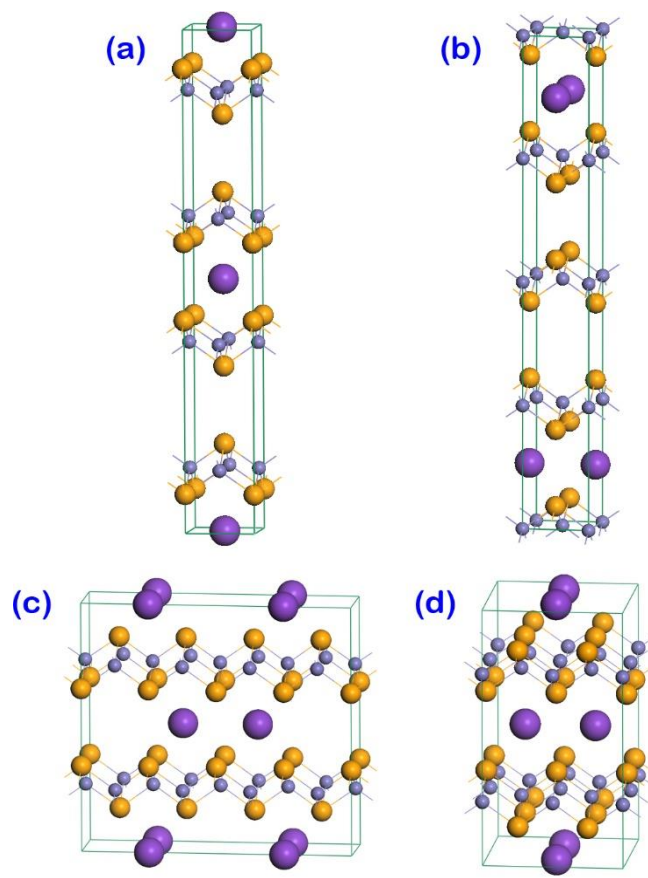

Figure S1 (a) (b) The structures of two kinds of $1 \times 1 \times 2$ supercells with composition of $K_{0.5}Fe_2Se_2$. The former structure always has a vacancy between two K ions while the K ions in the latter appear in pairs along $c$ axis. (c) The structure of $4 \times 1 \times 1$ supercell with composition of $K_{0.5}Fe_2Se_2$. (d) The structure of $2 \times 2 \times 1$ supercell with composition of $K_{0.5}Fe_2Se_2$. Violet (largest), grey (smallest) and orange spheres represent K, Fe and Se atoms, respectively.

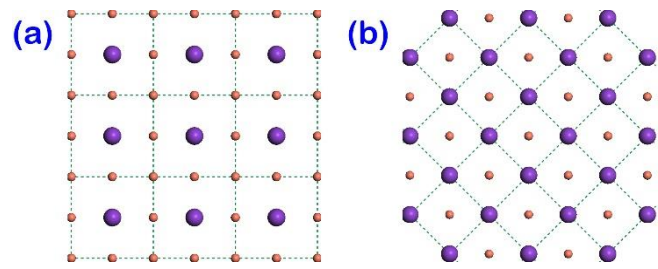

Figure S2 The possible K ion arrangements in $K_{0.25}Fe_2Se_2$ (a) and $K_{0.5}Fe_2Se_2$ (b). Violet big spheres represent K ions and orange small spheres represent vacancies.

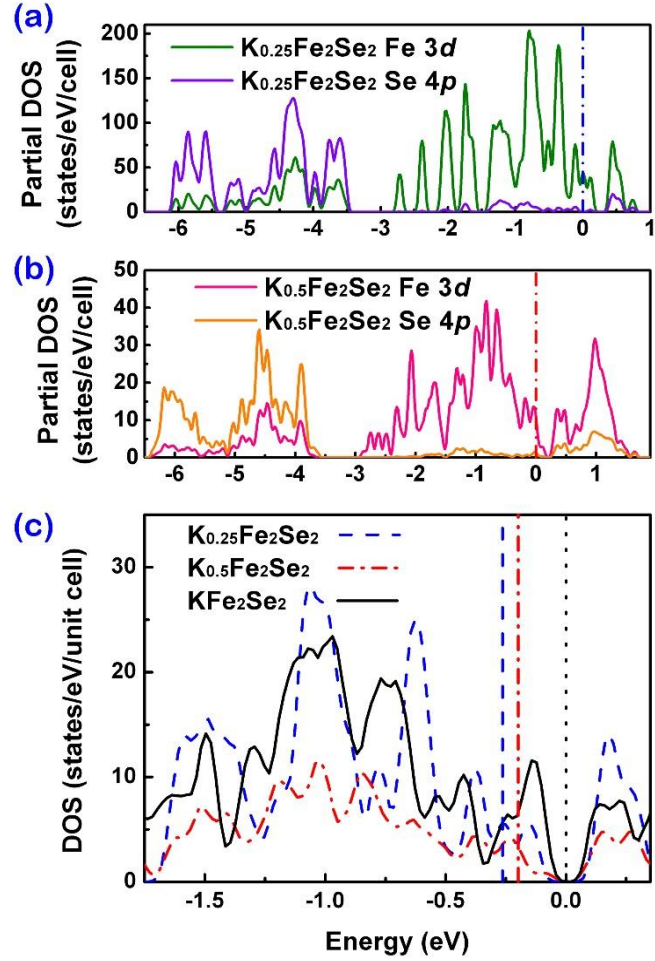

Figure S3 (a) (b) The partial DOS of Fe 3$d$ and Se 4$p$ in $K_{0.25}Fe_2Se_2$ and $K_{0.5}Fe_2Se_2$. The Fermi level is set to zero. (c) The comparison of total DOS details around the Fermi levels of $K_{0.25}Fe_2Se_2$, $K_{0.5}Fe_2Se_2$ and $KFe_2Se_2$. The Fermi level of $KFe_2Se_2$ is set to zero and others are set accordingly.